\documentclass[11pt]{article}
\usepackage{graphicx}
\usepackage{color}
\begin{document}

\title{
Production mechanism of hot nuclei in violent collisions 
in the Fermi energy domain.
}

\author{ 
M. Veselsky
\thanks{Phone: (979)-845-1411, fax: (979)-845-1899, e-mail: veselsky@comp.tamu.edu}
\thanks{On leave of absence from Institute of Physics, Slovak Academy of Sciences, Bratislava, 
Slovakia}
\thanks{Address after Apr 1, 2002: Institute of Physics, Slovak Academy of Sciences, Dubravska cesta 9, Bratislava, Slovakia, e-mail: fyzimarv@savba.sk}
\\
{\small Cyclotron Institute, Texas A\&M University, 
College Station, TX 77843.} 
}

\date{ }

\maketitle

\begin{abstract}
{
A production mechanism of highly excited nuclei formed in violent collisions 
in the Fermi energy domain is investigated. The collision of two nuclei 
is decomposed into several stages which are treated separately. 
Simplified exciton concept is used for the description of pre-equilibrium 
emission. A modified spectator-participant scenario is used 
where motion along classical Coulomb trajectories is assumed. 
The participant and one of the spectator zones undergo incomplete fusion. 
Excitation energies of both cold and hot fragment are determined. 
Results of the calculation are compared to recent experimental 
data in the Fermi energy domain. Data on hot projectile-like, mid-velocity 
and fusion-like sources are described consistently. 
Geometric aspects of pre-equilibrium emission are revealed. 
Explanations to previously unexplained experimental phenomena are given. 
Energy deposited into non-thermal degrees of freedom is estimated. \\
\\
{\it PACS:} 24.10.-i; 24.10.Lx; 25.70.-z; 25.70.Lm; 25.70.Mn; 25.70.Pq\\
{\it Keywords:} Nuclear reactions; E = 20 - 100 \hbox{A MeV}; 
Pre-equilibrium emission; Incomplete fusion; Multifragmentation; 
Collective flow
}
\end{abstract}

\section*{Introduction}

A detailed knowledge of the production mechanism of hot nuclei is desirable 
for understanding the processes leading to multifragmentation. 
The peripheral collisions of heavy ions in the Fermi energy domain 
demonstrate that nucleon exchange is a dominating mechanism 
leading to the production of highly excited quasiprojectiles \cite{TaGo,VePRC}. 
In more violent collisions, the processes leading to mid-velocity emission 
start to play an important role. The influences of pre-equilibrium 
emission and fragmentation-like processes become important. 
The hot source is created which further undergoes multifragmentation. 
The origin and properties of the hot mid-velocity source remain still 
a matter of discussions. In order to obtain properties of the hot source 
the experimental studies performed with large angular coverage 
( preferably in 4$\pi$-geometry ) are of primary interest. 
Considerable amounts of such data were obtained 
during last decade \cite{Dore,Toke,Frnk1,Frnk2,Dago1,Dago2}. 
In this article, a model of production of highly excited nuclei 
in violent collisions based on several simple phenomenological assumptions 
is presented and its capability to explain trends of the experimental 
observables related to dynamical properties and de-excitation 
of the hot source created in violent collisions is investigated. 

\section*{Model}

\begin{figure}[!htbp]
\centering
\vspace{5mm}
\includegraphics[width=5.cm,height=11.cm,angle=270]{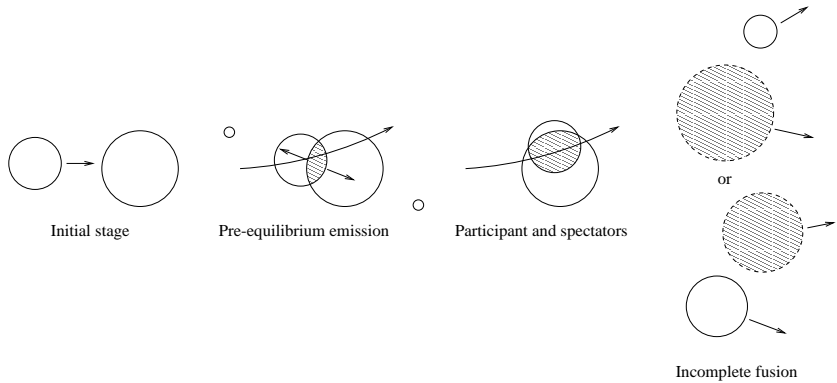}
\caption{ \footnotesize 
Schematic description of the physical picture employed in the 
model calculation for violent collisions. Shaded areas mean 
overlapping and later hot regions. 
}
\label{fgmod}
\end{figure}

The physical picture employed in the calculations is depicted in 
Fig. \ref{fgmod}. Properties of the highly excited nuclei in violent collisions 
are determined on an event-by-event basis in the Monte Carlo fashion. 
The model considers several stages of the collision. Different stages of the 
collision are treated separately. First, pre-equilibrium particle emission 
takes place. Later, the intermediate projectile-target system 
is reconstructed and the participant and spectator zones are determined. 
Finally, an incomplete fusion channel is chosen via interaction 
of spectators with the participant zone. 

\subsection*{Pre-equilibrium emission}

The pre-equilibrium emission ( PE ) is a process where fast particles are 
emitted prior to the equilibration of the system. The emission 
of fast pre-equilibrium particles in the reactions induced by nucleons 
and light particles was theoretically explained using 
the exciton model \cite{Griff}. For reactions induced by heavy-ion beams 
a model of nucleon exchange was developed \cite{RanVan}. In the present work, 
we use a phenomenological description \cite{VeZPA} based on similar assumptions 
as the exciton model. The probability of pre-equilibrium emission 
for a given reaction stage is evaluated using the formula 
\begin{equation}
\label{ppre}
P_{pre}(n/n_{eq})=1-e^{-\frac{(n/n_{eq}-1)^2}{2\sigma ^2}}
\end{equation}
for $n\le n_{eq}$ and equals zero for $n>n_{eq}$, 
where $n$ is the number of excitons at a given stage and $n_{eq}$ is the 
the number of excitons in the fully equilibrated compound system ( consisting 
of both projectile and target ) for a given 
excitation energy. The basic assumption leading to equation (\ref{ppre}) 
is that P$_{pre}$ depends exclusively on the 
ratio $n/n_{eq}$ as can be deduced from the results of ref. \cite{Boh} 
where the density of particle-hole states is approximately described 
using a Gaussian centered at $n_{eq}$. The parameter $\sigma$ is a free 
parameter of the calculation and no dependence on excitation energy is assumed. 
An initial exciton number is equal to the mass number of the projectile 
nucleus. The equilibrium 
number of excitons in the fully equilibrated compound system is calculated 
according to the formula \cite{Boh} 
\begin{equation}
\label{neq}
n_{eq}=2\hbox{ }g\hbox{ }T\hbox{ }\ln2
\end{equation}
where $g$ is the single particle level density at the Fermi energy and $T$ is 
the nuclear temperature determined as $T^2=U/\tilde{a}$, where $\tilde{a}$ is 
the level density parameter ( $\tilde{a}=A/9$ ) 
and $U$ is the excitation energy. At each 
emission step, a random number between zero and one is generated. If the random 
number is smaller than P$_{pre}$ a pre-equilibrium particle is emitted. 
If no pre-equilibrium emission occurs at a given emission stage, 
the pre-equilibrium stage is finished. 

The properties of emitted pre-equilibrium particles are determined 
for every emission. The type of particle emitted is determined randomly using 
the Hauser-Feschbach emission widths for neutron, proton and $\alpha$-particle. 
A Maxwellian spectrum of kinetic energy with an apparent 
temperature \cite{FuMo,GeBa} 
\begin{equation}
\label{tappa}
T_{app}= [\frac{2.5}{A_{P}}(E_{P}-V_{C})]^{1/2}
\end{equation}
is assumed, where A$_P$ is the projectile mass number, E$_P$ is the projectile 
energy and V$_C$ is the Coulomb barrier of projectile and target. 
The apparent temperature corresponds to the excitation energy of the 
fireball formed by the projectile nucleus together with another A$_P$ nucleons 
from the target nucleus. This model describes successfully 
experimental systematics of measured apparent temperatures as 
shown in refs \cite{FuMo,GeBa}. Particles are emitted isotropically from 
the fireball frame moving at half the projectile velocity. After emission, 
the exciton number is increased by a value 
\begin{equation}
\label{dltn}
\Delta n=A_{pre}\hbox{ }\frac{\kappa}{\beta_{0r}(l)}
\end{equation}
where A$_{pre}$ is the mass of emitted particle, $\beta_{0r}(l)$ is the 
radial velocity in the contact configuration at a given angular 
momentum and $\kappa$ is a free parameter. 

\subsection*{Projectile and target after pre-equilibrium emission}

Since pre-equilibrium emission is assumed to occur prior to the fragmentation 
stage, it is necessary to reconstruct the post-pre-equilibrium 
projectile-target configuration. The conclusions of the work 
\cite{RanVan} imply that in the asymmetric reactions with light projectile and 
heavy target the pre-equilibrium particles are mostly emitted from the 
projectile and propagate through the target. 
In the present work we assume that the net mass loss caused by emission 
of pre-equilibrium particles is distributed between projectile and target 
according to the relation 
\begin{equation}
\label{dmass}
\frac{\Delta A_{P}}{\Delta A_{T}}=\frac{A_{T}}{A_{P}}
\end{equation}
which complies to the conclusions of work \cite{RanVan} for asymmetric 
systems and gives equal net mass losses for the symmetric case. 
For the charge, a distribution of the net charge loss similar 
to Eqn. (\ref{dmass}) would lead to unbound projectile species 
for mass asymmetric projectile-target systems. To avoid that, a net 
charge loss of the projectile is chosen as 
\begin{equation}
\label{dchrg}
\Delta Z_{P}= \Delta A_{P} \frac{Z_{P}+Z_{T}}{A_{P}+A_{T}}.
\end{equation}
Since the emission from the fireball \cite{FuMo,GeBa} implies 
that an emitted particle undergoes typically one nucleon-nucleon collision, 
one can expect that the total excitation energy of the projectile and target 
after the pre-equilibrium stage would track with the sum of kinetic energies of 
the emitted pre-equilibrium particles. On the other hand, in the case 
when multiple pre-equilibrium particles were already emitted one should 
take into account the possibility that any further nucleon-nucleon collision 
can also lead to the decrease of the excitation energy. We employ a formula 
\begin{equation}
\label{exrw}
E^{*}_{tot} = ( 1 + \sqrt{\Delta A_{pre}} ) \frac{E_{k}^{pre}}{\Delta A_{pre}}
\end{equation}
where $E^{*}_{tot}$ is the total excitation energy, $E_{k}^{pre}$ is the sum 
of kinetic energies of all pre-equilibrium particles and $\Delta A_{pre}$ is 
the net mass loss due to pre-equilibrium emission. Formula (\ref{exrw}) is an 
approximation to the case where a random walk in excitation energy starts 
after the first two emissions. The excitation energy of the projectile 
and target is proportional to their masses. This is consistent 
with formula (\ref{dmass}) since every particle originating from 
the projectile propagates through the target \hbox{( and} vice versa ) 
where a collision occurs. 

\subsection*{Spectators and participant}

Several geometrical models of fragmentation have been proposed 
before where projectile and target are supposed to follow 
a straight trajectory determined by an impact parameter \cite{Gosset,Dayras}. 
Other works assume a classical Coulomb trajectory up to the closest approach 
configuration and only later the nuclei are supposed to follow a straight 
line \cite{Abul-Magd,Harvey}. 
In the Fermi energy domain, where angular momentum plays an important role, 
the classical Coulomb trajectories are more realistic since 
conservation of angular momentum is assured. 

In the present work, we assume a classical Coulomb trajectory of the 
intermediate projectile-target system without making any additional 
assumptions. The minimum distance between the intermediate projectile and 
target is used as a principal parameter of the geometric overlap scenario. 
For any possible trajectory, the overlapping volume is not  
smaller than the overlap of two spheres at minimum distance ( closest 
approach ). On the other hand, the geometric overlap formula of the 
abrasion-ablation model \cite{Gosset} always gives an overlap volume 
for a given minimum distance larger than the one along the classical Coulomb 
trajectory. The exact result is between these two values. 
We determine the volume of the participant zone randomly 
from the interval with limiting values given 
by the abrasion-ablation formula and the two-sphere overlap formula. 
Such a value differs from the exact result but nevertheless can be 
a reasonable approximation when taking into account irregularities 
of the separating nuclear surfaces. 
As a result, one participant and one or two spectator zones are created 
in the fragmentation stage. Their masses are determined proportionally 
to the determined volumes. 
The charges of the spectators are determined according to the 
combinatorial probability density \cite{Fried,Gaim} 
{
\begin{equation}
\label{ffried}
P(Z_{iS}) = \frac{(^{Z_{i}}_{Z_{iS}})(^{N_{i}}_{N_{iS}})}{(^{A_{i}}_{A_{iS}})}
\end{equation}
}
where 
A$_{i}$, Z$_{i}$, N$_{i}$ are the mass number, charge and neutron number 
of the projectile ( target ) 
and A$_{iS}$, Z$_{iS}$, N$_{iS}$ are the mass, charge and neutron 
number of the projectile ( target ) spectator.  

\subsection*{Incomplete fusion}

In the Fermi energy domain one can assume that the participant zone 
will not necessarily exist individually but can be captured by either 
the projectile or the target spectator zone. Especially in the symmetric 
reaction it is reasonable 
to assume that the capture by either of the spectators should be equally 
probable. To make a choice, for both spectators the volumes were determined
within a distance of 1 fm from the separation plane in order 
to estimate the number of spectator nucleon which interact 
with the participant nucleons via nuclear interaction. 
The volume was approximated by a 1 fm thick segment of the sphere 
touching the participant zone in the closest approach configuration. 
The number of neighboring nucleons ( $A_{NS}$ ) is then 
determined using a Gaussian distribution centered at the value exactly 
corresponding to the volume with the standard deviation equal 
to $\sqrt{A_{NS}}$. The participant zone is captured by a spectator with 
more neighboring nucleons. 
The capturing spectator and participant zone form a hot fragment. 
The remaining spectator zone is much colder. 

The excitation energy of the cold fragment is determined assuming that the 
part of the kinetic energy of the relative motion of the cold 
fragment and participant zone is transferred into the internal heat 
during the separation via collisions of the spectator 
and participant nucleons along the separation plane. 
The formula for excitation energy reads 
\begin{equation}
\label{xdiss}
E^{*}_S=x\hbox{ }A_{NS}\hbox{ }(\frac{E_{P}-V_{C}}{A_{P}})\hbox{ }
\frac{<s>}{\lambda}
\end{equation}
where $E_{P}$ and $A_{P}$ are the kinetic energy and the mass number 
of the intermediate projectile after pre-equilibrium emission, $V_{C}$ is the 
Coulomb barrier between the cold and hot fragment in the contact configuration 
and $<s>=\frac{8<r_{seg}>}{3\pi}$ is the mean path of the spectator nucleon 
within the touching segment of the sphere along the separation plane. For the 
mean free path $\lambda$ a value 6 fm is adopted. 
For each collision half of the asymptotic kinetic energy 
is converted into heat on average ( $x$ is a random number between 
zero and one ). 

The kinetic energy and the emission angle of the cold fragment are determined 
randomly using the double differential cross section formula based 
on the Serber approximation \cite{Matsu} 
\begin{equation}
\label{fmatsu}\frac{d^{2}\sigma}{dE_{a}d\Omega_{a}} = \frac {
(E_{a}E_{b})^{1/2} } { (2\mu B_{P} + 2m_{a}^{2}E_{P}/m_{P} + 2m_{a}E_{a} -
4(m_{a}^{3}E_{P}E_{a}/m_{P})^{1/2}cos\theta )^{2} }
\end{equation}
where it is the fragment $a$ which flies away and the fragment $b$ which fuses 
with the other nucleus, $E_{a}$ and $E_{b}$ are their c.m. kinetic energies, 
$ B_{P} $ is the binding energy of $a$ and $b$ in $P$, $\mu$ is the reduced 
mass of the system $a+b$ , $m_{P},m_{a},m_{b}$ are the masses of $P,a,b$ and 
$\theta$ is the emission angle of $a$ with respect to the direction of $P$ 
in the closest approach. 

In the case where the cold fragment originates from the target, the system is 
transformed into the inverse frame, formulas (\ref{xdiss}) and (\ref{fmatsu}) 
are used and the system is transformed back into the normal frame. 
The excitation energy, kinetic energy and the angle of a hot fragment 
are determined from the kinematics. 
The intrinsic angular momentum of the hot fragment 
is calculated using a mean radial distance and momentum of the 
participant zone relative to the capturing spectator in the closest 
approach configuration. An orbital angular momentum is determined 
from the relative motion of the cold and hot fragment in the contact 
configuration and the intrinsic angular momentum of the cold fragment 
is determined assuming conservation of the total angular momentum.  

\section*{Results and discussion}

In order to investigate the proposed model we performed 
an extensive comparison of the results of model calculation for different 
reaction stages to available ( mostly not explained satisfactorily 
by models used in original works ) data obtained 
in the Fermi energy domain. Since the experimental data usually contain 
the events originating 
from both peripheral and violent collisions, the calculation was 
carried out ( unless specified \hbox{otherwise )} for the angular momentum 
range from zero to grazing angular momentum. The number of events per partial 
wave was proportional to the angular momentum. 
No ad hoc criterium distinguishing between peripheral 
and violent collisions was implemented. Instead, for each event the Monte Carlo 
deep-inelastic transfer ( DIT ) code of Tassan-Got \cite{TaGo} was used 
after pre-equilibrium stage. 
There it is assumed that a di-nuclear configuration is created 
only when the overlap of nuclei does not exceed 3 fm. 
An excited quasi-projectile and quasi-target were created in such 
cases and de-excitation followed. When the overlap exceeded 3 fm, 
the collision was considered violent and a spectator-participant 
concept was implemented. For the reactions in inverse kinematics, 
namely when the projectile is heavier than the target, the system was 
transformed into the inverted frame where the projectile becomes a target 
and vice versa. Then the calculation proceeded as described above and 
the final kinematic properties of the reaction products were obtained 
after a transformation back into the laboratory frame. 

\subsection*{Multiplicities of pre-equilibrium particles}

The model of pre-equilibrium emission was compared to the results of 
work \cite{Agni}. There a multiplicity of the pre-equilibrium particles 
was determined in coincidence with the projectile-like fragments 
\hbox{( PLF )} in the reactions of Ca beams with $^{112}$Sn target 
at 35 \hbox{A MeV}. The measured multiplicities have been found 
significantly higher than predictions by nucleon exchange model 
( NEM ) \cite{RanVan}. 

\begin{figure}[!htbp]
\centering
\vspace{5mm}
\includegraphics[width=7.cm,height=6.cm]{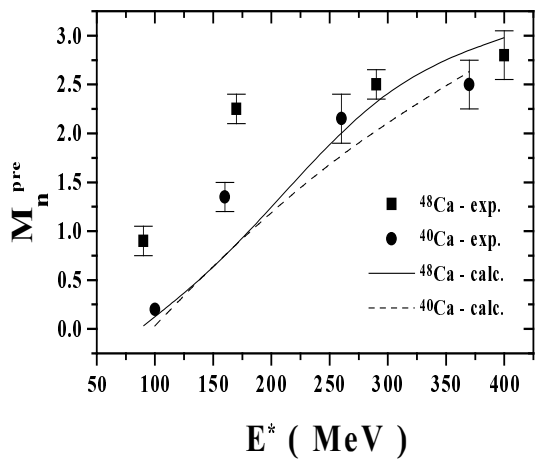}
\caption{ \footnotesize 
Experimental \cite{Agni} and calculated mean multiplicities 
of pre-equilibrium neutrons as a function of total excitation energy. 
Solid squares - experimental multiplicities 
measured in reaction of 35 \hbox{A MeV} $^{48}$Ca beam with $^{112}$Sn target, 
solid circles - ditto for 35 \hbox{A MeV} $^{40}$Ca beam, 
solid line -  calculated multiplicities in reaction 
of 35 \hbox{A MeV} $^{48}$Ca beam with $^{112}$Sn target, 
dashed line - ditto for 35 \hbox{A MeV} $^{40}$Ca beam. 
}
\label{fgagni}
\end{figure}

Fig. \ref{fgagni} gives the values of the pre-equilibrium neutron 
multiplicity in the reactions $^{40,48}$Ca+$^{112}$Sn for several bins of  
the total excitation energy. The solid squares and circles 
represent the results of work \cite{Agni} and the lines represent 
the results of the calculation.  
There are some differences for the bins with lower excitation energies 
( especially for $^{48}$Ca+$^{112}$Sn ) 
which can be caused by inconsistency of experimentally determined 
and calculated excitation energies for the peripheral collisions. 
For more central bins the agreement is quite good. The parameters 
$\sigma$=0.22 and $\kappa$=0.3 were used in the calculation. The same values 
of $\sigma$ and $\kappa$ were used in other reactions and lead to 
results which track well with the results of experimental works 
where multiplicities of pre-equilibrium particles were determined 
\hbox{( or} at least estimated ) in coincidence with the heavy residues 
or fission fragments \cite{Holub,Hilscher,VeZPA} 
or in coincidence with the reconstructed quasi-projectile \cite{VePRC}. 

\begin{figure}[!htbp]
\centering
\vspace{5mm}
\includegraphics[width=7.cm,height=6.cm]{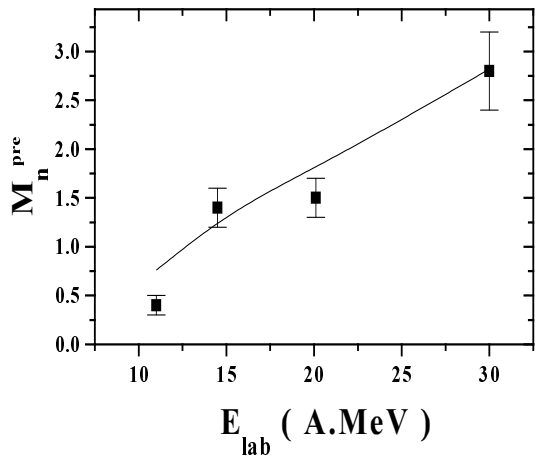}
\caption{ \footnotesize 
Experimental \cite{Holub,Hilscher} and calculated mean multiplicities 
of pre-equilibrium neutrons as a function of kinetic energy 
of the projectile. Solid squares - experimental multiplicities 
measured in reactions of 11-30 \hbox{A MeV} $^{20}$Ne beam 
with $^{165}$Ho target, 
solid line -  calculated neutron multiplicities for the same reaction. 
}
\label{fgholub}
\end{figure}

Fig. \ref{fgholub} gives multiplicities of pre-equilibrium neutrons 
for the reaction $^{20}$Ne+$^{165}$Ho at the projectile energies 
11 - 30 \hbox{A MeV} \cite{Holub,Hilscher}. The squares represent 
the experimental data and the line represents the calculation using parameters 
$\sigma$=0.22 and $\kappa$=0.3. The overall agreement is reasonable. 
Also for this data, the NEM model \cite{RanVan} was used and underestimated 
the experimental data ( a note added in proof in the original 
article \cite{RanVan} suggests 
possibility to improve the agreement between experiment and calculation ). 
The work \cite{VeZPA} estimates the probability value of 0.3 for the emission 
of four pre-equilibrium nucleons in coincidence with evaporation residues 
with Z$\ge$87 in the reaction $^{20}$Ne+$^{208}$Pb at projectile 
energy 15 \hbox{A MeV}. For the emission of six pre-equilibrium particles 
the probability value 0.1 is given. Corresponding calculated values 
( l$\le$50 ) are 0.25 and 0.07. Thus the calculation describes reasonably 
the multiplicity of pre-equilibrium particles in the central collisions 
at projectile energy 15 \hbox{A MeV}. 
The multiplicity of pre-equilibrium protons 
was estimated in the reaction $^{28}$Si+$^{112}$Sn for peripheral 
collisions in coincidence with the reconstructed hot projectile-like 
fragments \cite{VePRC}. The experimental estimates of proton 
multiplicities 0.2 for 30 \hbox{A MeV} and 0.3 for 50 \hbox{A MeV} correspond well 
to the calculated values 0.25 and 0.34, respectively. The calculated 
values were obtained using the model of pre-equilibrium stage, 
the DIT code \cite{TaGo} and the filtering procedure. 
The comparisons show that given description of multiplicities 
of the pre-equilibrium particles is consistent with the experimental trends. 
We adopt the parameters $\sigma$=0.22 and $\kappa$=0.3 as a standard set 
for further studies. 

Based on the comparisons of the calculation to experimental data 
one can try to understand the physical 
essence of the parameters $\sigma$ and $\kappa$ and to make conclusions 
concerning the nature of the process of pre-equilibrium emission. 
The parameter $\sigma$ can be, in principle, related to the width of 
the distribution of the particle-hole states which peaks at $n_{eq}$ 
and can be approximated by a Gaussian. The value $\sigma$=0.22 is 
about three time larger than the variance of the distribution 
of particle-hole states calculated using the formula given in \cite{Boh} for 
the range of the masses and excitation energies corresponding to the reactions 
considered here. Since the assumptions made in \cite{Boh} are valid 
only for moderate excitations, a larger value of $\sigma$ suggests 
that either the distribution of the particle-hole states at high excitations 
is wider than the distribution calculated using formula from \cite{Boh}  
or a larger value of $\sigma$ means that $n_{eq}$ 
grows slower with excitation energy than predicted in \cite{Boh}. 
Compared to the value $\sigma$=0.3 in the original paper \cite{VeZPA} 
the value $\sigma$=0.22 means decrease of approximately 30 \%. 
The increase in the probability of pre-equilibrium emission is 
counterbalanced by angular momentum profile of $\Delta n$ causing 
decrease of pre-equilibrium multiplicity at large angular momenta ( in the 
original paper no angular momentum dependence of $\Delta n$ was assumed ). 
The value of $\kappa$=0.3 suggests that, in the given range of radial 
velocities, the exciton number increases typically by 2 - 4 excitons 
between two subsequent pre-equilibrium emissions. 
This implies that particle emissions follow each other rather quickly since 
the exciton number increase between emissions is comparable 
with the minimum step of the exciton model ( $\Delta n$=2 ) 
towards equilibrium.    

\begin{figure}[!htbp]
\centering
\vspace{5mm}
\includegraphics[width=7.cm,height=6.cm]{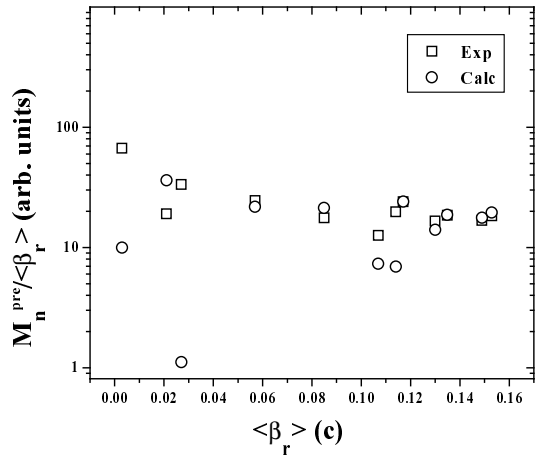}
\caption{ \footnotesize 
A dependence of the ratio of the mean multiplicity of pre-equilibrium 
neutrons to the mean radial velocity as a function of the latter. 
Experimental \cite{Agni,Holub,Hilscher} ( squares ) 
and calculated ( circles ) multiplicities 
have been used. 
}
\label{fgmnbrad}
\end{figure}

As follows from the relation (\ref{dltn}) the average rate of emitted 
pre-equilibrium particles per exciton number step during the collision 
( $1/\Delta n$ ) can be considered proportional to the radial velocity in the 
contact configuration $\beta_{0r}(l)$ for a given angular momentum. Since 
$\beta_{0r}(l)$ is an initial radial velocity it would be more 
appropriate to relate the emission rate to a mean radial velocity during 
the interaction, which can be expected to track with $\beta_{0r}(l)$. 
Fig. \ref{fgmnbrad} shows a dependence of the ratio of the mean multiplicity 
of pre-equilibrium neutrons to the mean radial velocity as a function of the 
latter. The mean radial velocity was calculated for a given angular momentum 
range along the classical Coulomb trajectories from the contact configuration 
to the point where $\beta_{r}$ drops under the threshold value 4 \hbox{A MeV}. 
The data from the Fig. \ref{fgagni} and \ref{fgholub} are used in 
Fig. \ref{fgmnbrad}. For the data from Fig. \ref{fgagni}, 
the calculated angular momentum ranges corresponding to the excitation 
energy bins were used. For the data from Fig. \ref{fgholub}, 
the angular momentum range from zero to grazing angular momentum was used. 
As one can see in Fig. \ref{fgmnbrad}, the ratio $M_{n}^{pre}/<\beta_{r}>$ 
is practically constant for experimental ( squares ) and with 
few exceptions also for calculated \hbox{( circles )} multiplicities. 
Thus, the mean multiplicity of pre-equilibrium neutrons ( and particles 
in general ) is proportional to a mean radial velocity $<\beta_{r}>$ 
determined for a corresponding range of angular momenta along 
the classical Coulomb trajectories. 

\begin{figure}[!htbp]
\centering
\vspace{5mm}
\includegraphics[width=7.cm,height=6.cm]{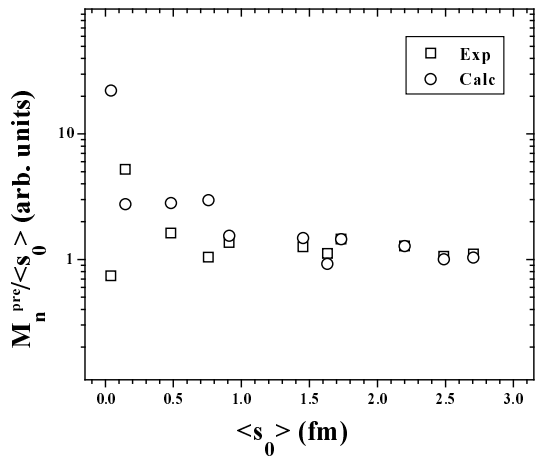}
\caption{ \footnotesize 
A dependence of the ratio of the mean multiplicity of pre-equilibrium 
neutrons to the mean overlap $M_{n}^{pre}/<s_0>$ as a function 
of $<s_0>$. Experimental \cite{Agni,Holub,Hilscher} ( squares ) 
and calculated ( circles ) multiplicities have been used. 
}
\label{fgmnover}
\end{figure}

The multiplicity of pre-equilibrium 
particles for a given collision can be expressed as radial integral 

\begin{equation}
\label{mpre}
M^{pre} = \int dr \frac{dM^{pre}}{dn} \frac{dn}{dr} ,
\end{equation}

where $r$ is the distance from the center of the target nucleus at a given 
point along the classical Coulomb trajectory. 
The exciton number change per radial distance $\frac{dn}{dr}$ can 
be related to the nucleon-nucleon cross section via $\frac{dn}{dl}$ which 
represents an exciton number change per unit distance along the trajectory. 
When assuming that the number of excitons increases as a result of 
two-body collisions, the observable $\frac{dn}{dl}$ can be directly related to 
the nucleon-nucleon cross section and can be considered proportional 
to $1/\beta$. Then $\frac{dn}{dr}$ becomes proportional to $1/\beta_{r}$. 
The validity of the relation (\ref{dltn}) can be extended to 
any moment of the reaction 
and the emission rate $\frac{dM^{pre}}{dn}$ can be considered proportional 
to $\beta_{r}$. Then, the integrand in (\ref{mpre}) becomes a constant and 
the mean multiplicity 
of pre-equilibrium particles should be also proportional to the mean value 
of radial overlap $<s_0>$ where radial velocity drops below the threshold 
value 4 \hbox{A MeV} which can be expected to track with $<\beta_{r}>$. 
In Fig. \ref{fgmnover} we show a dependence of the ratio of multiplicity 
of pre-equilibrium neutrons to the mean overlap $M_{n}^{pre}/<s_0>$ 
as a function of $<s_0>$. Indeed, one can see that for the mean overlap 
larger than 0.8 fm the ratio is again practically constant for both 
experimental ( squares ) and calculated \hbox{( circles} ) multiplicities 
( experimental values are practically constant from 0.3 fm \hbox{above ).} 
Thus, the picture given by equation (\ref{mpre}) can be considered 
consistent with the essential features of the process. 

When making conclusions concerning Figs. \ref{fgmnbrad} and 
\ref{fgmnover}, caution is necessary since no deceleration by 
the recoil from the emitted particles is considered. Nevertheless, 
we assume that even then the picture will not change dramatically. 
Then, taking into account the comparison of experimental data on 
multiplicities of the pre-equilibrium particles with the results 
of model calculations and the results of following geometric analysis, 
the process of pre-equilibrium emission can be qualitatively understood 
as a process of two-body dissipation along the classical Coulomb trajectory 
where the nucleons scattered in radial direction are emitted. In this picture, 
the pre-equilibrium stage stops when the radial motion disappears 
and the relative motion of the projectile and target is mostly tangential. 
When the tangential motion is slow enough, friction force can transform it 
into rotation of a di-nuclear system as in deep-inelastic collisions. 
When the tangential motion is fast, a violent collision follows and 
formation of the participant and spectator zones along the classical Coulomb 
trajectory appears to be a natural next stage of the collision. 

\subsection*{Projectile-like fragments}

In the recent experimental work \cite{Casini} a linear correlation 
between the primary mass of the projectile-like fragment and 
the net mass loss due to the de-excitation was reported in the 
nearly symmetric reactions of $^{93}$Nb with $^{116}$Sn at 25 \hbox{A MeV} 
in both normal and inverse kinematics for different dissipation bins. 
The net mass loss increases with the primary mass of the projectile-like 
fragment. With increasing dissipation this trend occurs in the still 
broader range of primary masses. Since the net mass loss is correlated 
to the excitation energy of the hot primary projectile-like nucleus, 
one can expect a similar trend for the excitation energy. It was 
demonstrated that both rapid growth of mass variance and asymmetric 
excitation energy sharing can not be explained within the concept 
of nucleon exchange through window between nuclei ( the NEM code 
\cite{RanDIT} was used in the analysis ). As an explanation, 
a dynamical scenario as a neck rupture was suggested.  

\begin{figure}[!htbp]
\centering
\vspace{5mm}
\includegraphics[width=7.cm,height=8.cm]{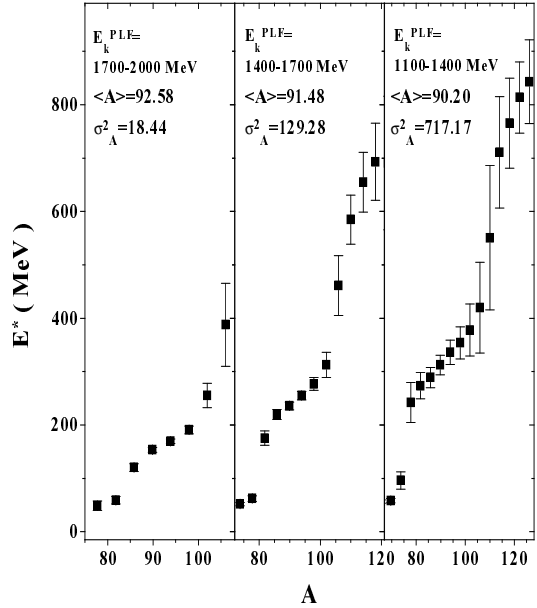}
\caption{ \footnotesize 
Calculated correlation between the primary mass and the excitation energy of 
a projectile-like fragment in the reaction $^{93}$Nb with $^{116}$Sn 
at 25 \hbox{A MeV}.
}
\label{fgcasi}
\end{figure}

Fig. \ref{fgcasi} gives a calculated correlation between the excitation 
energy and the mass of the hot projectile-like nucleus for different bins of 
kinetic energy. The calculation mimiques the experimental trend of the 
emitted mass ( a measure of excitation energy ) 
shown in Fig. 10 of the experimental article 
\cite{Casini}. At masses close to the beam the deep-inelastic transfer 
takes place but the range of primary masses is limited. To achieve 
larger mass changes more violent collision should occur. When the target 
strips a part of the projectile, the projectile-like fragment remains 
relatively cold. Hot projectile-like fragments are produced if a part 
of the target is picked-up by the projectile. 
For DIT events the mean excitation energy per nucleon is practically constant 
what leads to a slight increase of excitation energy with increasing mass. 
At the region where the violent collisions 
start to dominate a rapid change in excitation energy per nucleon 
takes place. The calculated correlations deviate from straight lines in the 
transition regions between deep-inelastic and incomplete fusion scenario. 
This discrepancy can be caused by a sharp cutoff value of the overlap 
implemented in the DIT code. It could be possibly improved by employing 
a diffuse cutoff but this is beyond the scope of this article. 
Also in Fig. \ref{fgcasi} are given mean masses and mass variances 
of the projectile-like fragment for given bins of kinetic energy. 
While the mean mass remains roughly constant the mass variance 
grows rapidly. The values of mass variance $\sigma_A^2$ are comparable 
to the values given in Fig. 19 of the experimental paper \cite{Casini}. 
The incomplete fusion appears to describe correctly both large 
mass variances and asymmetric excitation energy sharing and 
offers viable explanation of anomalies reported in \cite{Casini}. 

\begin{figure}[!htbp]
\centering
\vspace{5mm}
\includegraphics[width=8.cm,height=6.cm]{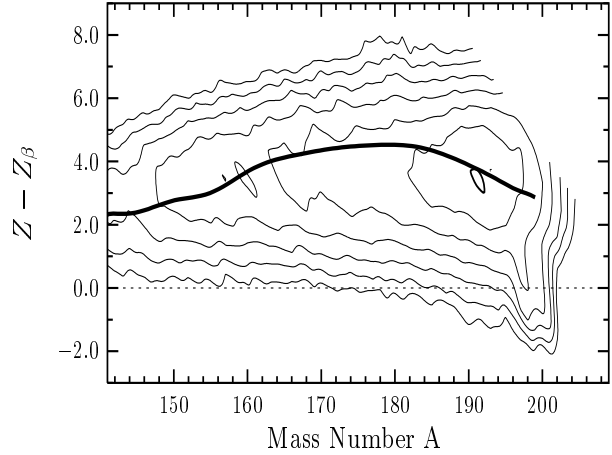}
\caption{ \footnotesize 
Measured yields \cite{Sou} of heavy residues at the forward angles 
in the reaction $^{197}$Au(20 \hbox{A MeV})+$^{nat}$Ti as a function of A and Z. 
Z is expressed relative to the line of $\beta$-stability. 
Solid line - calculated centroids of the fragment charge for given residue 
mass ( the code GEMINI \cite{GEM} was used for \hbox{de-excitation ).} 
}
\label{fgsoul1}
\end{figure}
 
The production of heavy residues in the reaction $^{197}$Au+$^{nat}$Ti 
was measured recently in the inverse kinematics by Souliotis et al. \cite{Sou} 
at the projectile energy 20 \hbox{A MeV}. Fig. \ref{fgsoul1} shows the measured 
yields of heavy evaporation residues at the forward angles as a function 
of A and Z ( Z is represented relative to the line of $\beta$-stability ). 
The experimental work \cite{Sou} gives comparison to relativistic fragmentation 
code \cite{EPAX} which reproduces reasonably the charge centroids 
for masses far from the beam even if the validity of such approach 
at 20 \hbox{A MeV} is questionable. The solid line in Fig. \ref{fgsoul1} 
represents the centroids of Z for a given residue mass calculated in the 
present work. The angle and momentum cuts corresponding to the angle 
and momentum acceptance of the spectrometer were applied and 
the statistical code GEMINI \cite{GEM} was used for the de-excitation stage. 
One can see that the calculated centroids 
follow the experimental trend quite well. Thus it offers an explanation 
of the data appropriate for this projectile energy. 

\begin{figure}[!htbp]
\centering
\vspace{5mm}
\includegraphics[width=8.cm,height=6.cm]{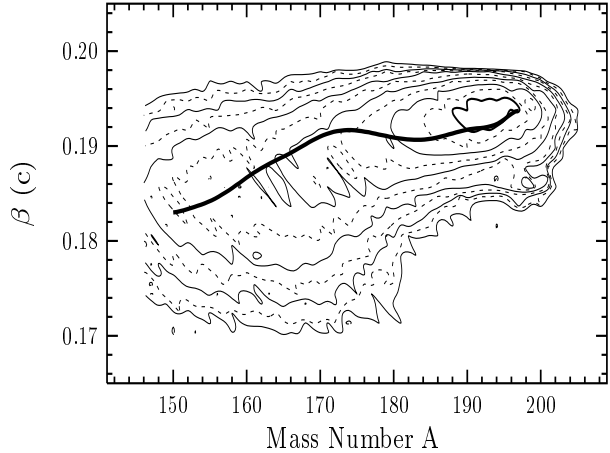}
\caption{ \footnotesize 
Experimental velocity distributions \cite{Sou} of the heavy residues 
measured at the forward angles 
in the reaction $^{197}$Au(20 \hbox{A MeV})+$^{nat}$Ti plotted as a function of A. 
Solid line - calculated velocity centroids for given residue mass.} 
\label{fgsoul2}
\end{figure}

A dominant mechanism contributing to the production of heavy residues 
can be understood from Fig. \ref{fgsoul2} where the measured velocity 
distributions are given as a function of fragment mass. With decreasing 
mass the reaction products become still slower. Again, the line represents 
the calculated centroids after the spectrometer cut. One can see that 
the calculation follows the experimental data. A typical contributing scenario 
in the calculation is the emission of up to three 
pre-equilibrium neutrons followed by deep-inelastic transfer and by 
de-excitation of the heavy fragment ( the excitation energies do not exceed 
3 \hbox{A MeV} ). A similar agreement was obtained for the evaporation 
residues in the more symmetric reaction $^{197}$Au(20 \hbox{A MeV})+$^{90}$Zr 
\cite{Sou2}. The calculation seems to be able to provide adequate treatment 
of the pre-equilibrium stage and offers a natural extension of the original 
DIT code of Tassan-Got \cite{TaGo}. 

\subsection*{Reaction dynamics and properties of the hot source}

\begin{figure}[!htbp]
\centering
\vspace{5mm}
\includegraphics[width=7.cm,height=8.cm]{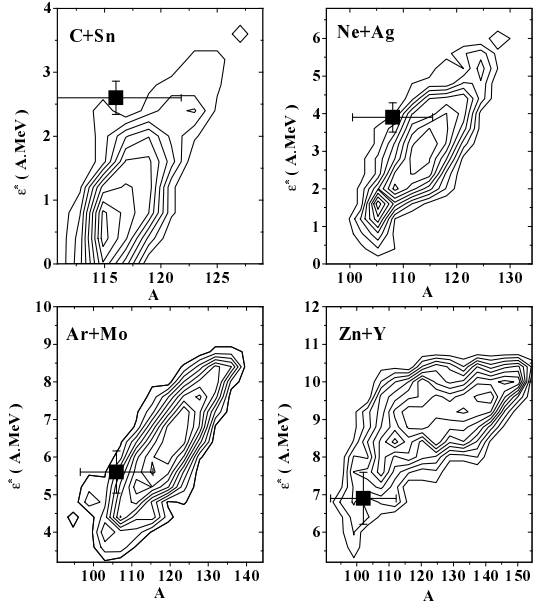}
\caption{ \footnotesize 
Calculated correlation of the excitation energy and the mass of the hot 
target-like sources produced in the reactions $^{12}$C+$^{116}$Sn, 
$^{20}$Ne+$^{108}$Ag, $^{40}$Ar+$^{100}$Mo and $^{64}$Zn+$^{89}$Y 
at projectile energy 47 \hbox{A MeV}. Solid squares show 
values from \cite{Hagel}. 
}
\label{fgkris}
\end{figure}

The mass and excitation energy of the heavy fusion-like source have 
been determined in the work \cite{Hagel} ( the former indirectly using 
coalescence analysis and QMD simulation \cite{Chim} and the latter 
using three-source fit and light charged particle ( LCP ) calorimetry 
techniques ) in four projectile-target 
combinations with the same projectile energy 47 \hbox{A MeV}. The experiment 
provided practically full coverage for LCPs at angles beyond 20$^{\circ}$. 
Fig. \ref{fgkris} gives the calculated distributions of excitation energy 
versus mass of the heavy source for reactions $^{12}$C+$^{116}$Sn, 
$^{20}$Ne+$^{108}$Ag, $^{40}$Ar+$^{100}$Mo and $^{64}$Zn+$^{89}$Y 
at the projectile energy 47 \hbox{A MeV}. Only the most central events 
have been included into calculation. The limiting angular momentum 
was set so that the partial waves included represent 10 \% of 
the reaction cross section. Black squares in all plots 
correspond to the data from \cite{Hagel}. In the case of the 
reaction $^{12}$C+$^{116}$Sn the target-like nuclei with the excitation energy 
not exceeding 2 \hbox{A MeV} do not contribute to the experimental data 
because of the low multiplicity of charged particles and the data  
point can be related only to the part of the calculated distribution above 
2 \hbox{A MeV}. For the reaction $^{20}$Ne+$^{108}$Ag the data  
point is located slightly away from the central part of the calculated 
distribution but both are still roughly consistent within experimental 
uncertainties. In the reactions $^{40}$Ar+$^{100}$Mo and 
$^{64}$Zn+$^{89}$Y the data  point is shifted from 
the center of the calculated distributions towards lower 
excitation energies and masses. This effect seems to increase 
with increasing symmetry of the projectile-target combination. 
For the reaction $^{40}$Ar+$^{100}$Mo the data  point is moved 
slightly towards lower mass and excitation energy but still remains 
close to the center of distribution. For the reaction $^{64}$Zn+$^{89}$Y 
the shift is already significant. 
A possible explanation 
of the shift can be the presence of an additional effect not 
included in the calculation such as collective flow which would lead 
to hardening of the spectra of transverse energies and thus to 
overestimation of direct emission within the three-source fit method 
and to underestimation of source excitation energy. Due to geometric coverage 
the three-source analysis as a model dependent technique was performed 
at angles 20$^{\circ}$ and larger in the laboratory frame which 
for the increasingly symmetric reactions correspond to even larger angles 
in the center of mass frame and an increasing amount of information from 
forward angles is missing. 
This puts less constraint onto the properties of projectile-like source 
in the fitting procedure and the emission multiplicity can be overestimated. 
More detailed data of the same type including also the forward angles 
would be of interest for further study.

\begin{figure}[!htbp]
\centering
\vspace{5mm}
\includegraphics[width=7.cm,height=8.cm]{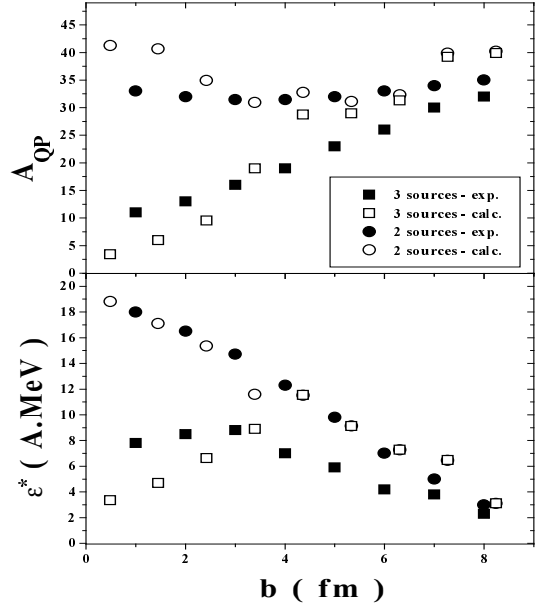}
\caption{ \footnotesize 
Experimental \cite{Dore} and calculated masses and excitation energies of 
the projectile-like source as a function of the impact parameter 
for the reaction $^{40}$Ar+$^{58}$Ni at 95 \hbox{A MeV}. 
Solid squares - experiment, 3 sources,  
solid circles - experiment, 2 sources,  
open squares - calculation, 3 sources,  
open circles - calculation, 2 sources.   
}
\label{fgdore}
\end{figure}

The reaction $^{40}$Ar+$^{58}$Ni at 95 \hbox{A MeV} was studied 
experimentally using the INDRA 4$\pi$ setup \cite{Dore}. The properties 
of the projectile-like source have been reconstructed using 
two different types of analysis. In the two-source analysis, 
the projectile-like source consisted of all the detected particles 
with parallel velocity larger than the center of mass velocity. 
The three-source analysis employs the usual three-source fit technique. 
Experimental masses and excitation energies of the projectile-like 
source are given in Fig. \ref{fgdore} as a function of the impact parameter 
( solid \hbox{ squares} - 3 sources, solid circles - 2 sources ). 
Open symbols represent the results 
of the calculations. As an equivalent of the three-source analysis 
the properties of the calculated projectile-like source are shown 
( open squares ). The two-source analysis was approximated by assuming 
Gaussian shapes of the both projectile- and target-like source with 
the width of 17 \% of the projectile rapidity as obtained in 
the experimental work \cite{Dore}. The properties 
of the projectile-like source ( open \hbox{circles )} have been obtained by 
an integration of the relative part of the target- and projectile-like source 
at velocities above the center of mass velocity. The calculated dependencies 
appear to follow the experimental ones reasonably well \hbox{( with} some 
discrepancies in the transition from peripheral to violent collisions similar 
to \hbox{Fig. \ref{fgcasi} ).} In the two-source analysis, the mass of the hot 
source is slightly overpredicted for most central collisions while obtaining 
excellent agreement in excitation energy. The properties obtained for 
the projectile-like source in the most central collisions can be identified 
with the properties of the forward half of the very hot composite mid-velocity 
source created by an incomplete fusion of the participant and spectator zone.  
The velocity of the composite source comes close to the center 
of mass velocity. In the three-source analysis, the mass of the 
projectile-like source in the most central collisions is underestimated 
but the experimental trend is reproduced better than using the geometrical 
calculation \cite{Dayras} presented in the experimental paper. The excitation 
energy in most central collisions is again underestimated. In this case, 
the experimental mass of about 10 with excitation energy 8 \hbox{A MeV} 
implies that mostly nucleons and light charged particles are observed. 
The question is to what extent the most energetic pre-equilibrium particles 
contribute to this source.

\begin{figure}[!htbp]
\centering
\vspace{5mm}
\includegraphics[width=7.cm,height=7.cm]{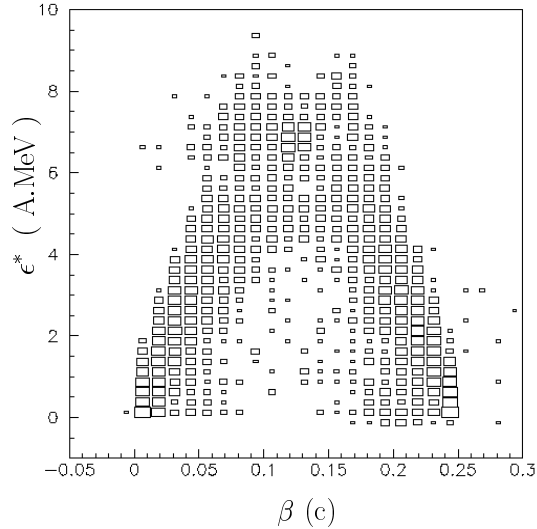}
\caption{ \footnotesize 
Calculated correlation of the excitation energy and velocity of the hot 
nuclei produced in the reaction $^{58}$Ni+$^{58}$Ni at 30 \hbox{A MeV}. 
}
\label{fgnini}
\end{figure}

The measurement of the production of intermediate mass fragments ( IMFs )
in symmetric collisions $^{58}$Fe,$^{58}$Ni+$^{58}$Fe,$^{58}$Ni 
at 30 \hbox{A MeV} 
\cite{Ramak} determined three different sources of IMFs, the moderately excited 
projectile(target)-like source at velocities close to the projectile (target) 
and the highly excited source at velocities close to the center of mass 
velocity. Fig. \ref{fgnini} shows a calculated correlation between the 
excitation energy and the velocity of the hot source in the laboratory frame 
calculated for the reaction $^{58}$Ni+$^{58}$Ni at 30 \hbox{A MeV}. 
Both projectile-like and target-like nuclei are included in the plot. One can  
identify three sources analogous to the ones seen in the experiment in the same 
reaction. The projectile- and target-like sources are moderately excited. 
The third source with the average velocity close to the mid-velocity 
is highly excited. Both the projectile- and target-like nuclei contribute 
to this source. 

\begin{figure}[!htbp]
\centering
\vspace{5mm}
\includegraphics[width=7.cm,height=7.cm]{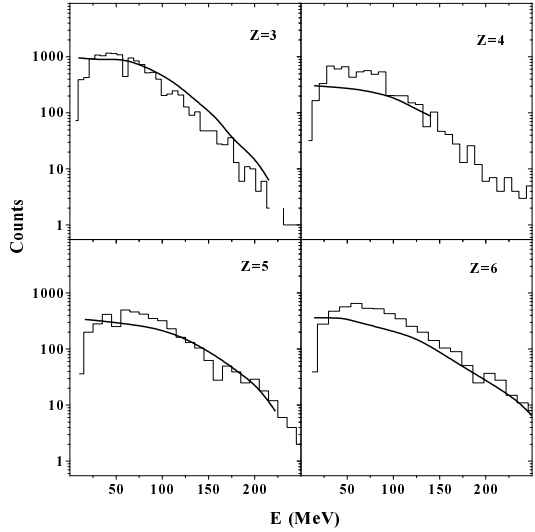}
\caption{ \footnotesize 
Kinetic energy spectra of emitted fragments with Z = 3 - 6 
at laboratory angle 40$^{\circ}$ for the 
reaction $^{58}$Fe+$^{58}$Ni at 30 \hbox{A MeV}. 
Experimental \cite{Ramak} spectra are depicted by thick solid 
lines approximating the experimental data points, calculation 
is represented by histograms. 
}
\label{fgnini2}
\end{figure}

In order to study this reaction in more detail, the de-excitation of hot 
fragments was simulated using the statistical SMM code \cite{SMM}. 
Both projectile- and target-like source have been de-excited. Each source was 
de-excited separately and no influence of Coulomb field of the other 
fragment was assumed. As one can see on Fig. \ref{fgnini2} the calculation 
is able to reproduce the experimentally observed fragment energy spectra 
at laboratory angle 40$^{\circ}$ for the same reaction. 
Experimental spectra are depicted by thick solid 
lines approximating the experimental data points, calculation 
is represented by histograms. The normalization was chosen in order 
to reproduce approximately the shape and the sum of the spectra for Z = 3 
at energies above 25 MeV. The shapes of spectra are reproduced satisfactory. 
Compared to QMD-SMM simulation the present calculation appears to reproduce 
better the low energy part below 50 MeV while performing comparably 
in the high energy part but this may be to some extent just the effect 
of different normalizations. 

\begin{figure}[!htbp]
\centering
\vspace{5mm}
\includegraphics[width=6.cm,height=7.cm]{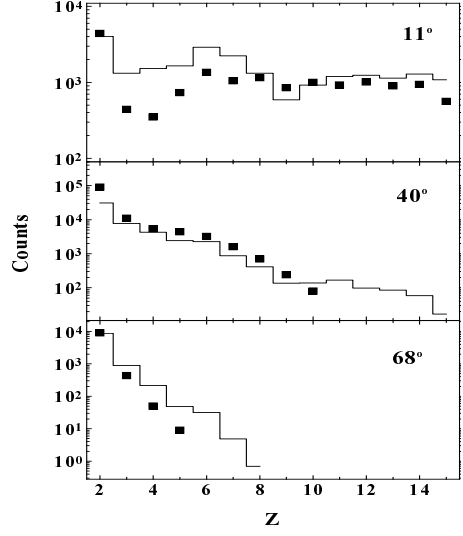}
\caption{ \footnotesize 
Fragment charge distributions at different angles for the 
reaction $^{58}$Fe+$^{58}$Ni at 30 \hbox{A MeV}. 
Solid squares give the experimental \cite{Ramak} charge distributions 
at 11$^{\circ}$, 40$^{\circ}$ and 68$^{\circ}$ and histograms represent 
the calculation.
}
\label{fgnini3}
\end{figure}

In Fig. \ref{fgnini3} the fragment charge distributions at different angles 
are given. Solid squares give the experimental charge distributions 
at 11$^{\circ}$, 40$^{\circ}$ and 68$^{\circ}$ and histograms represent 
the calculation. Calculated distributions have been filtered using 
the energy range of the Si-Si telescopes used. Furthermore, correction for 
solid angle coverage of detectors was made. As one can see the 
calculation again performs reasonably well. At 11$^{\circ}$ the 
experimental charge distribution is reproduced well for Z = 8 - 15. 
For Z = 3 - 7 calculation the experimental yields are overpredicted. 
According to the experimental paper \cite{Ramak}, for the same angle 
the QMD-SMM calculation underpredicted yields 
at Z = 8 - 15 while still overpredicting yields at Z = 3 - 7. Another 
calculation used in the experimental paper BUU-SMM largely overpredicted 
yields of lighter fragments while still underpredicting the yields 
of heavier fragments. Thus at forward angles the present calculation 
leads to better agreement than the calculations given in the original work. 
The reduction of the yields of fragments with Z = 3 - 7 can be possibly 
explained by an influence of Coulomb field of the heavy projectile-like 
fragment. At larger angles the present calculation also performs comparably 
( QMD-SMM ) or better than calculations given in the experimental paper. 

\begin{figure}[!htbp]
\centering
\vspace{5mm}
\includegraphics[width=6.cm,height=9.cm]{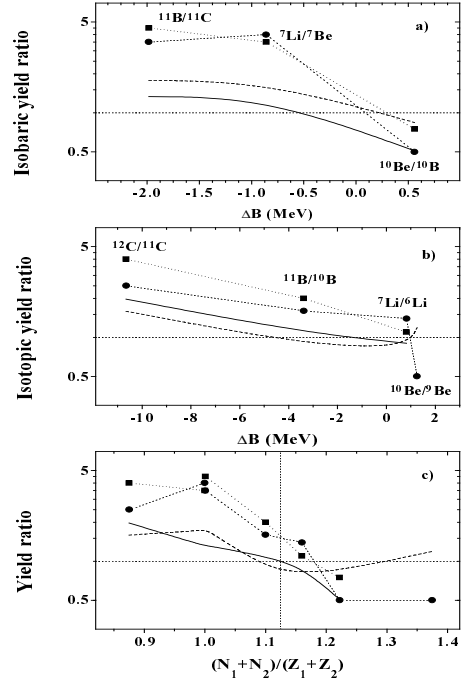}
\caption{ \footnotesize 
(a),(b) - Experimental \cite{Ramak} and calculated isobaric and isotopic yield 
ratios for the reaction $^{58}$Fe+$^{58}$Ni at 30 \hbox{A MeV} as a function 
of the difference of binding energies ( squares and circles - 
experimental values at 11$^{\circ}$ and 40$^{\circ}$ respectively, 
solid and dashed line - calculated values at 11$^{\circ}$ and 
40$^{\circ}$ respectively, horizontal dotted line - unity ). 
(c) - Fragment yield ratios from (a),(b) plotted against average N/Z-ratio 
of the fragment pair ( symbols and lines analogous to (a),(b), vertical dotted 
line shows average N/Z-ratio of stable isotopes with Z = 3 - 6 ). 
}
\label{fgnini4}
\end{figure}

\begin{figure}[!htbp]
\centering
\vspace{5mm}
\includegraphics[width=9.cm,height=12.cm]{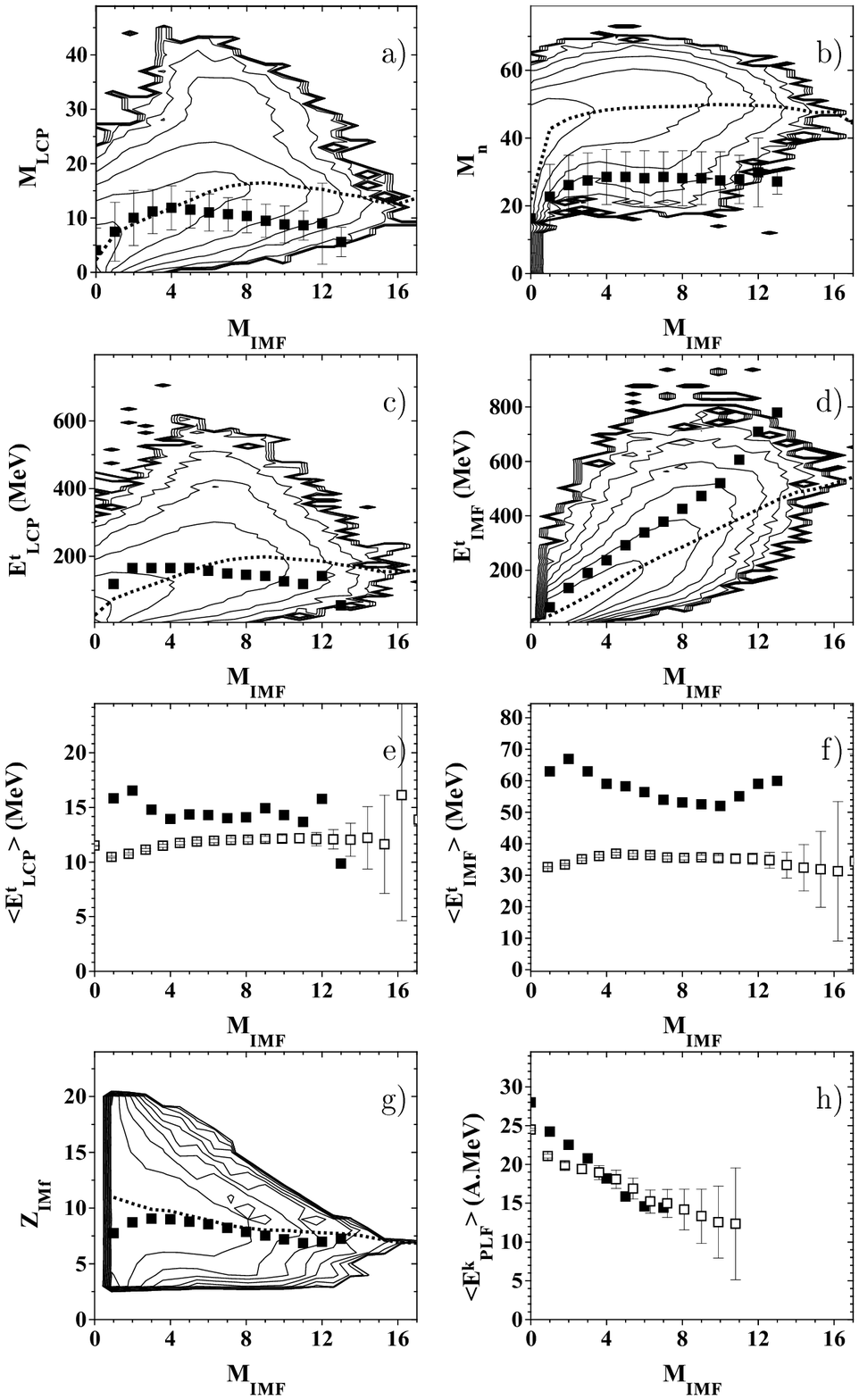}
\caption{ \footnotesize 
Comparison of measured observables \cite{Toke} to the results of calculations 
for the reaction $^{136}$Xe+$^{209}$Bi at 28 \hbox{A MeV}. 
Solid symbols - experimental mean values for given IMF multiplicity, 
open symbols or thick dashed lines - calculated mean values, 
contour plots - calculated distributions.   
(a),(b) - dependencies of LCP and neutron multiplicity on IMF multiplicity, 
(c),(d) - total transverse energies of LCPs and IMFs, 
(e),(f) - mean transverse energies of LCPs and IMFs, 
(g) - IMF charge distributions, (h) - PLF kinetic energy. 
}
\label{tkfig1}
\end{figure}

Furthermore, the experimental work \cite{Ramak} gives measured 
values of several isobaric and isotopic ratios at different angles. 
In Fig. \ref{fgnini4}a,b are given the values of experimental
and calculated isobaric and isotopic yield ratios for the reaction 
$^{58}$Fe+$^{58}$Ni at 30 \hbox{A MeV} as a function 
of the difference of binding energies ( squares and circles - 
experimental values at 11$^{\circ}$ and 40$^{\circ}$ respectively, 
solid and dashed line - calculated values at 11$^{\circ}$ and 
40$^{\circ}$ respectively, horizontal dotted line - unity ). 
As one can see the experimental yield 
ratios vary in much wider range than the calculated ones. The calculation 
typically underpredicts the yield ratios where yields of stable isotopes 
are compared to the yields of proton-rich isotopes. On the other hand, 
the yield ratios comparing the yield of neutron-rich isotope $^{10}$Be 
to the yields of less neutron-rich isotopes are overpredicted. This situation 
is demonstrated clearly in Fig. \ref{fgnini4}c where both isobaric 
and isotopic yield ratios are plotted against average N/Z-ratio 
of a pair of nuclei ( the vertical dotted line shows average 
N/Z-ratio of the stable isotopes with Z = 3 - 6 ). Thus, the production 
of both proton- and neutron-rich isotopes seems to be overpredicted 
in the calculation when compared to the yields of stable isotopes. 
Similar situation was observed also in the original work when 
using both QMD-SMM and BUU-SMM calculations. 
Since the slope of plots in Fig. \ref{fgnini4}a,b can be related 
to temperature, it seems that the isotopic composition of fragments 
is determined at lower temperature than it is assumed in the SMM 
calculation. A possible explanation can be the collective flow of particles 
induced in the most violent phase of the collision. The inconsistencies 
in description of de-excitation in the SMM code can be possibly ruled 
out since it performs reasonably well in the case where hot quasiprojectiles 
have been created in the peripheral collisions of $^{28}$Si beam with 
$^{112,124}$Sn targets at 30 and 50 \hbox{A MeV} \cite{VePRC,isodist}. 
There the mechanism of quasiprojectile production is described well 
using DIT-code of Tassan-Got \cite{TaGo} while the SMM code \cite{SMM} 
describes reasonably well the de-excitation. The simulation employing both 
codes gives good description of the isobaric 
ratio Y($^{3}$H)/Y($^{3}$He) \cite{isodist}. 
In such a reaction the collective flow apparently does not play a role. 
In the case of reaction $^{58}$Fe+$^{58}$Ni at 30 \hbox{A MeV} 
one can attribute the inconsistencies in description of isotopic and 
isobaric ratios at central angles to non-thermal effects rather than 
to a breakdown of description of de-excitation by SMM code at transition from 
masses 20-30 to 60-90. 
The kinetic energy of a collective flow can not 
be included into thermal energy and the temperature at the freeze-out 
becomes lower. In any case, the isotopic distributions seem to provide 
detailed insight into reaction dynamics. The calculation presented here  
is able to describe the inclusive characteristics like spectra and 
charge distributions at different angles and provide correct overall 
characteristics of the hot source. Nevertheless, additional assumptions 
concerning the amount of energy transformed into thermal degrees of freedom 
are necessary ( the same applies also to QMD-SMM and BUU-SMM calculations 
presented in the original work \cite{Ramak} ). Further studies are necessary 
in this direction. 

In a recent experimental work Toke et al. \cite{Toke} reported 
the observation of a new dynamical mechanism of fragment emission. 
The reaction $^{136}$Xe+$^{209}$Bi at 28 \hbox{A MeV} was studied in 
virtually 4$\pi$-geometry.  
Several experimental observables have been studied as a function 
of the multiplicity of intermediate mass fragments with Z = 3 - 20. 
Independence of the charge and transverse energy of IMFs, 
binary character of the reaction, simultaneous saturation of the multiplicity 
of neutrons and light charged particles ( saturation of heat content ) and 
absence of the competition between IMF emission and thermal emission 
have been interpreted as the main evidences of a new process where 
intermediate mass fragments are emitted dynamically without competition 
with emission of light particles. The IMFs in the events with 
IMF multiplicity 2 and 4 were emitted dominantly at parallel velocities 
between 0 and 5 cm/ns in the c.m. frame.  

The experimental results have been compared to the results of 
the calculation. As in the previous case the SMM code was used 
for de-excitation. Both sources have been de-excited independently. 
Results  are given in the Fig. \ref{tkfig1}. For realistic comparison 
the calculated IMF multiplicities have been corrected to the geometric 
coverage of the detector setup which was 90 \% \cite{Toke2}. 
As one can see the calculation reproduces the simultaneous 
saturation of neutron and LCP multiplicity ( Fig. \ref{tkfig1}a,b ). 
The absolute values of multiplicity in the saturation region are overestimated 
for LCPs even when taking into account geometric coverage of the experimental 
setup ( in the case of LCPs the overall detection efficiency can be 
estimated to be close to total geometrical coverage ). 
In the case of neutrons the detection efficiency was estimated 
to be close to 70 \% \cite{Toke2} what brings the calculated 
neutron multiplicities just to the upper experimental limit. 
The trends of total transverse 
energies are tracked reasonably well for both LCPs and IMFs 
( Fig. \ref{tkfig1}c,d ). The experimental transverse energies 
of both IMFs and LCPs are underpredicted by the calculation by 
a practically constant amount of energy for a wide range of IMF multiplicities 
( Fig. \ref{tkfig1}e,f ) and the experimental IMF charge and energy 
of projectile-like fragment are reproduced well for IMF multiplicities 
above 4 which can be attributed to violent collisions ( Fig. \ref{tkfig1}g,h ). 
For lower IMF multiplicities where peripheral collisions dominate there 
are inconsistencies ( Fig. \ref{tkfig1}e,f,g,h ) which can be attributed 
in part to the inconsistencies in the transition region 
observed also in other reactions. Also the geometric restrictions of 
the experimental setup at forward angles not taken into account in the 
calculations can be a source of inconsistencies. The increase of the 
IMF transverse energy at highest experimental multiplicities 
( Fig. \ref{tkfig1}d,f ) is based on data points with very low statistics. 

\begin{figure}[!htbp]
\centering
\vspace{5mm}
\includegraphics[width=9.cm,height=12.cm]{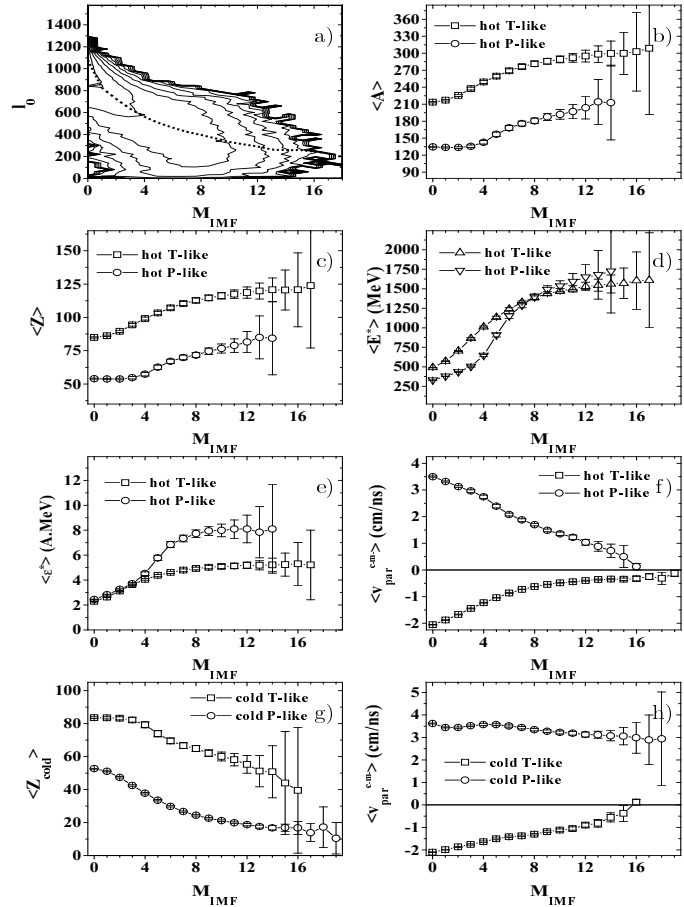}
\caption{ \footnotesize 
Calculated observables for the reaction $^{136}$Xe+$^{209}$Bi 
at 28 \hbox{A MeV}. 
(a) - distribution of initial angular momenta ( thick dashed line - 
mean values for given IMF multiplicity ), (b) - (h) - 
various calculated characteristics 
of both projectile- and target-like hot and cold sources.  
}
\label{tkfig2}
\end{figure}

As one can see on Fig. \ref{tkfig1} the calculation 
reproduces the overall trends of the experimental data which 
have been interpreted as an evidence of dynamical emission 
of IMFs. The effect which was observed experimentally appears 
to be caused by a gradual increase of the mass and excitation energy 
of the emitting source with increasing centrality. 
Various calculated characteristics 
of both projectile- and target-like hot and cold sources are 
given in Fig. \ref{tkfig2}. As one can see the mass and charge 
of the hot target-like source ( which is produced in violent collisions 
with the probability exceeding 90 \% ) reach values up to 300 
( Fig. \ref{tkfig2}b ) and 120 ( Fig. \ref{tkfig2}c ), respectively, 
while excitation energy reaches value 5 \hbox{A MeV} ( Fig. \ref{tkfig2}d ). 
For such a source, relatively small change in the mass, charge and 
excitation energy can lead to opening or closing of various 
emission channels. For example, the inconsistency in the LCP 
saturation multiplicity observed in Fig. \ref{tkfig1}a 
can be corrected by subtracting 0.5 \hbox{A MeV} from the 
excitation energy. This, in analogy to previous reactions 
suggests that the nominal excitation energy of the hot source is not fully 
transferred into thermal excitation energy which determines the properties 
of the fragment partition at the freeze-out. Part of the source 
energy is transferred into kinetic energy of collective motion resulting 
in the non-thermal flow. In the particular 
case of the reaction $^{136}$Xe+$^{209}$Bi at 28 \hbox{A MeV} 
the collective flow energy 
can be roughly estimated by above mentioned value 0.5 \hbox{A MeV}. 
Such a non-thermal flow is also suggested by observed excess of experimental 
transverse energy of LCPs and IMFs when compared to calculation 
( Fig. \ref{tkfig1}e,f ). There the collective flow energy can be estimated 
to approximately 3 MeV per fragment charge unit leading to 
the value of about 1.2 \hbox{A MeV} for transverse direction. 
Thus the collective flow energy of LCPs and IMFs in the transverse direction 
appears to be larger than the overall decrease of excitation energy due to 
flow. It is of interest for further studies to investigate if it 
is caused by anisotropic profile of collective flow energy or if it coincides 
with lowering of transverse energy of other reaction products. 

Parallel velocity of the emitted fragments can be related 
to the calculated parallel velocities of both hot and cold 
projectile- and target-like sources ( Fig. \ref{tkfig2}f,h ). 
In the dominant scenario the hot target-like source moves 
at parallel velocity slightly below c.m. velocity while 
the cold projectile-like source has parallel velocity 
3 - 4 cm/ns. According to the experimental paper \cite{Toke} the 
IMFs in mostly peripheral collisions with IMF multiplicity 2 and 
4 are emitted predominantly in the forward direction in the c.m. frame. 
Calculated parallel velocities for such collisions imply that the IMF angular 
distribution is strongly influenced by the Coulomb field of the cold fragment 
( Fig. \ref{tkfig2}g ) which causes IMFs to be emitted with 
highest probability in the configuration 
between hot and cold source which has the lowest Coulomb energy. 
Such interaction can be the cause of higher 
experimental LCP and IMF transverse energies in peripheral collisions 
( Fig. \ref{tkfig1}e,f ). 

\begin{figure}[!htbp]
\centering
\vspace{5mm}
\includegraphics[width=7.cm,height=6.cm]{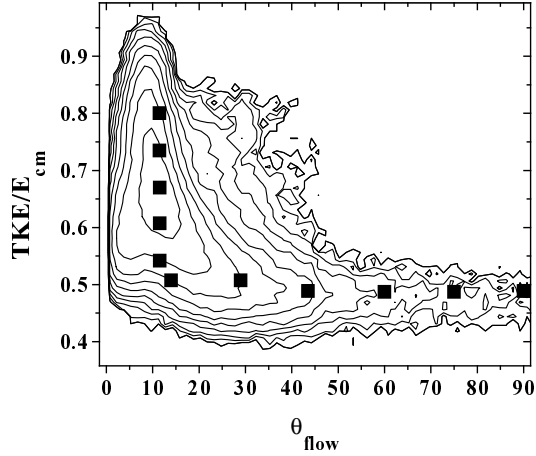}
\caption{ \footnotesize 
Total kinetic energy plotted versus flow angle $\theta_{flow}$ 
for the reaction $^{155}$Gd+$^{nat}$U at 36 \hbox{A MeV}. 
Calculated TKE/$\theta_{flow}$-plot is shown as 
contour plot while solid squares show experimental points \cite{Frnk1}
with $\Omega$ = 0., 0.1, ... , 1.0. Total kinetic energy is 
expressed relative to the available c.m. energy. 
}
\label{frnkfig1}
\end{figure}

In the very recent experimental works \cite{Frnk1,Frnk2} 
the reaction $^{155}$Gd+$^{nat}$U was studied at projectile 
energy 36 \hbox{A MeV}. The experiment was carried out with 
geometric coverage close to 4$\pi$. In order to select the most 
central single-source events, a detailed event shape analysis 
was carried out. The events have been classified using the 
plot of total kinetic energy ( TKE ) vs. flow angle $\theta_{flow}$. 
Flow angle was calculated for each event using its sphericity and coplanarity. 
An observable $\Omega$ with values from 0. to 1. was defined 
in the most populated regions of the plot as a measure 
of centrality of the event. 
In the present work, the experimental TKE/$\theta_{flow}$-plot 
was compared to the results of model calculation. As in previous 
cases, the SMM code \cite{SMM} was used as afterburner and 
both sources have been de-excited independently. For each 
event a shape analysis was performed, sphericity and coplanarity 
were calculated and the flow angle was determined. Calculated 
TKE/$\theta_{flow}$-plot is shown in Fig. \ref{frnkfig1} as 
a contour plot while solid squares give experimental points 
with $\Omega$ = 0., 0.1, ... , 1.0. Total kinetic energy is 
expressed relative to the available c.m. energy. As one can see the 
calculation is consistent with the experimental points. This allows 
to obtain dependencies of various calculated observables 
on $\Omega$. 

\begin{figure}[!htbp]
\centering
\vspace{5mm}
\includegraphics[width=7.cm,height=6.cm]{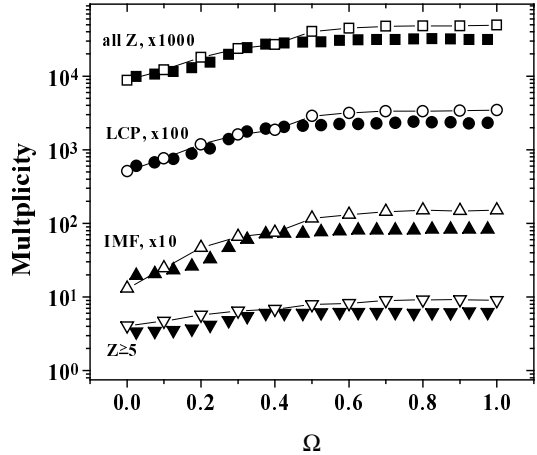}
\caption{ \footnotesize 
Multiplicities of various reaction products plotted as a function 
of $\Omega$ for the reaction $^{155}$Gd+$^{nat}$U at 36 \hbox{A MeV}. 
Solid symbols - measured multiplicities \cite{Frnk1} 
of all charged particles ( squares ), 
light charged particles ( circles ), 
intermediate mass fragments ( up triangles ) 
and fragments with Z $\ge$ 5 ( down triangles ). 
Open symbols - calculated multiplicities for corresponding subsets 
of reaction products.
}
\label{frnkfig2}
\end{figure}

In Fig. \ref{frnkfig2} are given multiplicities of various 
reaction products as a function of $\Omega$. Solid symbols represent 
the measured multiplicities of all charged particles, light charged 
particles, intermediate mass fragments and fragments with Z $\ge$ 5. 
Open symbols show calculated quantities. Both experimental and 
calculated multiplicities exhibit saturation for $\Omega \ge$ 0.5. 
Calculated saturation multiplicities typically overestimate 
experimental values. 

In a similar way to the reaction $^{136}$Xe+$^{209}$Bi 
the inconsistency in multiplicities can be corrected by removing part of the 
excitation energy. In the present case in the most central 
collisions the calculated mass, charge and excitation energy 
of the hot source are A = 386, Z = 153 and $\epsilon^{*}$ = 6.35 \hbox{A MeV}. 
In this case, both target- and projectile-like source contribute 
because their excitation energy exceeds 3.5 \hbox{A MeV} what 
makes the event look like single-source event. Such events have been 
experimentally observed and their mass, charge and excitation energy 
have been determined using the LCP calorimetry method \cite{Frnk2} 
( A$_{exp}$ = 378, Z$_{exp}$ = 150 
and $\epsilon^{*}$$_{exp}$ = 6.5 \hbox{A MeV} ). 
Using the calculated properties of the single-source the observed 
saturation multiplicities can be reproduced with excitation energy 
being decreased by 1 \hbox{A MeV} ( detection efficiency of about 
90 \% was assumed ). This can be considered as 
an estimate of the energy of the collective flow which is consistent 
with estimate from experimental work 0.5 $\pm$ 0.5 \hbox{A MeV} obtained 
using the c.m. fragment kinetic energy spectra. 

\begin{figure}[!htbp]
\centering
\vspace{5mm}
\includegraphics[width=7.cm,height=6.cm]{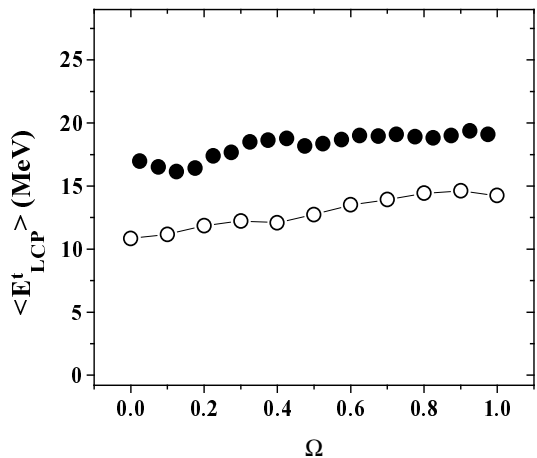}
\caption{ \footnotesize 
Mean transverse energy of light charged particles depending on $\Omega$ 
for the reaction $^{155}$Gd+$^{nat}$U at 36 \hbox{A MeV}. 
Solid symbols - experiment \cite{Frnk1}, open symbols - calculation. 
}
\label{frnkfig3}
\end{figure}

Further estimate of the collective flow energy can be obtained 
from mean transverse energy of light charged particles which is 
shown in Fig. \ref{frnkfig3} as a function of $\Omega$. In analogy 
to the reaction $^{136}$Xe+$^{209}$Bi the calculation underestimates 
the mean values of transverse energy by the same amount ( close to 
5 MeV ) for a wide range of events with different centrality. 
The kinetic energy of the collective flow can be in this case 
estimated as 3.5 MeV per charge leading to flow energy 
approximately 1.4 \hbox{A MeV}. Thus, again in analogy to 
the reaction $^{136}$Xe+$^{209}$Bi, the energy of the 
collective flow of the light charged particles in the transverse direction 
is larger than the value of excitation energy to be subtracted 
in order to reproduce experimental multiplicities. 

The experimental work \cite{Frnk2} further provides the 
multiplicities of pre-equilibrium ( direct ) particles 
emitted in coincidence with single-source events. For the 
events with $\theta_{flow}$ $>$ 70$^{\circ}$ typically 7 
pre-equilibrium emissions occurred ( 3 neutrons and 4 LCPs ). 
In the calculation, mean pre-equilibrium multiplicity 
accompanying single-source events is 6.5$\pm$0.5 
( M$_{n}^{pre}$=3.8$\pm$0.4 and M$_{LCP}^{pre}$=2.7$\pm$0.3 ).
There is reasonable agreement in overall multiplicity 
of pre-equilibrium emissions while the larger multiplicity 
of neutrons in the calculation can be possibly explained 
by more LCP emission channels in the experimental data ( p, d, t, $\alpha$ ) 
than in the calculation ( p, $\alpha$ ). 

Another heavy system studied recently in 4$\pi$-geometry was 
$^{197}$Au+$^{197}$Au at projectile energy 35 \hbox{A MeV} 
\cite{Dago1,Dago2}. For this symmetric heavy system 
an analysis of central collisions was performed with emphasis put 
on the properties of emitted fragments. For charge distributions 
measured in the most central collisions a model analysis was carried 
out using SMM code and mass, charge, excitation energy and flow energy 
of the single-source have been determined. Two sets of source parameters 
have been determined \cite{Dago2} which reproduced the data equally well - 
A = 343, Z = 138, $\epsilon^{*}$ = 6.0 \hbox{A MeV} ( flow energy 
was 0. and freeze-out density was set $\rho_{0}$/3 ) and  
A = 315, Z = 126, $\epsilon^{*}$ = 4.8 \hbox{A MeV} ( flow energy 
was 0.8 \hbox{A MeV} and freeze-out density was set $\rho_{0}$/6 ).  
Calculation analogous to previous cases gives for most central 
collisions ( l$_{0}$ $\le$ 100 ) a hot source with 
mean parameters A = 341, Z = 136, $\epsilon^{*}$ = 6.45 \hbox{A MeV}. 
Corresponding cold source still exists ( A = 52, Z = 22, E$^{*}$ = 110 MeV ) 
but de-excites mostly by neutron emission. The properties of the 
calculated source are in reasonable agreement with the heavier 
of the sources. In a similar way to previous reactions an estimate 
of thermal excitation energy at freeze-out was obtained using mean IMF 
multiplicity. In the most central collision of $^{197}$Au+$^{197}$Au 
at 35 \hbox{A MeV} the experimentally determined mean IMF multiplicity 
was 10.8$\pm$1. Calculation reproduces this value when subtracting 
0.75 \hbox{A MeV} ( detection efficiency of 90 \% was assumed ). 
Such a value of collective non-thermal energy 
is comparable with previous cases. The difference in the amount 
of collective flow between SMM calculations used in \cite{Dago2} 
and here can originate in the freeze-out hypothesis used. 
The SMM calculation used in \cite{Dago2} used a fixed freeze-out density 
that leads to fixed Coulomb barrier and higher sensitivity of 
mean c.m. kinetic energy to freeze-out temperature. 
In the present work the freeze-out density is determined 
according to prescription depending on fragment multiplicity \cite{SMM} 
in which freeze-out density ( and thus Coulomb barrier ) decrease 
with increasing fragment number thus leading to weaker sensitivity. 
The experimental mean fragment c.m. kinetic energies for most 
central collisions given in \cite{Dago2} can not be used 
in order to distinguish between the two hypotheses because of 
large uncertainties caused by low experimental statistics. 

In summary, the investigations presented in this section demonstrate 
that the model calculation provides consistent description of the 
production mechanism of the hot source which further undergoes 
multifragmentation. The calculation described the overall properties 
of the hot source like mass, charge and available c.m. energy 
for wide range of experimental data, especially well for the 
highly complete and model independent data obtained in 4$\pi$-geometry. 
Thermodynamical properties of the source determined from fragment 
partitions suggest that the energy available in the source frame is not 
fully transferred into heat and a collective flow starts to play role. 
This is indicated indirectly by several observables such as 
particle multiplicities, transverse energies and isotopic/isobaric ratios. 
In any case, the systematic trend rather than any particular case 
can be considered as an indication of the non-thermal flow. 
Further detailed studies on that subject possibly by investigating 
large set of observables in the same system would be of great interest. The 
estimated energy of the collective flow does not seem to differ dramatically 
for various systems and the values extracted are in good agreement 
with other studies ( e.g. \cite{Flow} and references therein ). 
Furthermore, the collective flow energy seems to 
be practically constant for events with different centralities. 
Such a behavior can be possibly explained within the present physical picture 
by the interplay of two different effects. In central collisions, 
the participant zone increases rapidly and compressional effects 
should lead to a flow concentrated in transverse direction. 
In peripheral collisions, the charge of the cold spectator 
is significant enough to focus the fragments emitted in the 
mid-velocity region into transverse direction. 
Such an enhancement of fragment multiplicity at central angles 
was described e.g. in work \cite{Botv}.

\subsection*{Possible limitations of the model}

With increasing projectile 
energy, the production of the spectator-participant-like three-body events 
should start to play an important role. As primary candidates for three-body 
events can be considered the events where the relative motion of the 
participant zone and the capturing spectator leads to values of intrinsic 
angular momenta above the critical angular momentum for fusion. 
Furthermore, at projectile energies much above the Fermi energy 
the intra-nuclear cascade should take place and change the properties 
of the pre-equilibrium source. For projectile energies below 20 \hbox{A MeV} 
the nucleus-nucleus interaction becomes more complex because of the 
effect of proximity potential. The sum-rule model \cite{WilczSum} 
of Wilczynski et al. employing a concept of angular momentum windows 
determined by critical angular momenta for different incomplete fusion 
channels can be used in this energy region.

\section*{Conclusions}

The comparisons to the wide range of experimental observables measured 
in various reactions in the Fermi energy domain appear 
to imply that the present approach describes consistently the main features 
of the violent processes leading to the production of excited projectile-like, 
mid-velocity and fusion-like sources in the projectile energy range between 
20 and 100 \hbox{A MeV}. Different stages of the collision can be 
distinguished and related to each other using a simple phenomenological 
assumptions based on a geometrical picture of the collision. 
The mean multiplicity of pre-equilibrium particles is described 
satisfactory by using phenomenological approach employing an exciton 
concept. Furthermore, a direct relation of the mean multiplicity 
of pre-equilibrium particles to the geometrical aspects of the reaction 
is demonstrated. The pre-equilibrium emission appears to be 
a consequence of the radial motion of the projectile-target system 
and ceases when radial motion is transferred into tangential 
via Coulomb interaction. At that point, according to the degree of 
tangential motion a deep-inelastic transfer or a formation of the participant 
and spectator zones follows. The participant zone can be captured by 
one of the spectator zones and incomplete fusion occurs. 
The Coulomb interaction between the projectile and target 
as a whole plays still an important role in such a violent 
reaction scenario. The resulting motion 
along the classical Coulomb trajectories furthermore 
assures conservation of angular momentum. 
Concerning the applicability of the model, it appears to 
describe consistently in many projectile-target systems 
the mass, charge and excitation energy 
of hot nuclei which later undergo multifragmentation. 
Furthermore, the presence of non-thermal 
effects can be expected from observed multiplicities, 
transverse energies and isotopic/isobaric yield ratios. 
The values of flow energy are similar for different 
systems and are practically constant within one projectile-target system 
for collisions with different centrality. 

The author would like to thank G.A. Souliotis and S.J. Yennello 
for the support and fruitful and stimulating discussions. Furthermore, 
the author would like to thank A. Sanzhur for discussions 
concerning pre-equilibrium emission. 
The author would also like to thank F.-P. Hessberger, D.D. Bogdanov, 
B.I. Pustylnik and A.V. Yeremin for many valuable discussions 
on low energy reaction mechanisms which helped to formulate 
some of the basic concepts of this work. Finally, the author 
would like to thank L. Tassan-Got for the use of his DIT code, 
R.J. Charity for the use of the code GEMINI and 
A.S. Botvina for the use of the SMM code. This work was supported in part by 
the Robert A. Welch Foundation through grant No. A-1266, 
the Department of Energy through grant No. DE-FG03-93ER40773, 
and through grant VEGA-2/1132/21.

\end{document}